\documentclass[prl,twocolumn,showpacs,amsmath,amssymb]{revtex4}

\usepackage{graphicx}

\usepackage{amsmath}
\begin{document}
\title{Analytic Constructions of General $n$-Qubit Controlled Gates}
\author{Yang Liu$^{1}$}
\author{Gui Lu Long$^{1}$\footnote{Corresponding author}}
\email{gllong@mail.tsinghua.edu.cn}
\author{Yang Sun$^2$}
\affiliation{$^1$Key Laboratory of Atomic and Molecular NanoSciencs and Department of Physics, Tsinghua University, Beijing 100084, P. R. China\\
$^2$ Department of Physics and Astronomy, University of Nortre Dame,
Indiana 46556, USA}
\date{\today}

\begin{abstract}
In this Letter, we present two analytic expressions that most
generally simulate $n$-qubit controlled-$U$ gates with standard
one-qubit gates and CNOT gates  using exponential and polynomial
complexity respectively. Explicit circuits and general expressions
of decomposition are derived. The exact numbers of basic operations
in these two schemes  are given using gate counting technique.
\end{abstract}

 \maketitle

\maketitle \emph{I. Introduction}-- The study of quantum computers
has been developing very rapidly over the past years. It provides
exponential speedup in factoring \cite{Shor}, or square-root speedup
in unsorted database search \cite{Grover}. In the circuit model of
universal quantum computer \cite{Deutsch}, the unitary operation
that completes a computation task is a series of gates on a fixed
number of qubits. Any unitary gate can be constructed from a set of
universal gates \cite{Deutsch,Lloyd}. Using the smallest number of
basic gates to construct an arbitrary unitary transformation is very
important, not only  for using less executing time, but also for
resulting less errors.

Complexity of circuit is measured in terms of the number of basic
gates, namely the one-bit gate and the two-bit CNOT gate. For a
general $2^n\times 2^n$ unitary matrix $U$ with $4^n$ degrees of
freedom, \emph{O}($4^nn^2$) elementary operations are needed in
principle \cite{Nielsen}.  Later on, efficient schemes implementing
arbitrary quantum gates have reduced the circuit complexity to
\emph{O}($4^n$) \cite{Vartiainen, Mottonen, Tucci}. They are
achieved by using the QR decomposition \cite{Vartiainen}, or the
cos-sin decomposition \cite{Mottonen}. General scheme for
decomposing an arbitrary gate is given in Ref. \cite{Tucci}  using
numerical method. For some quantum information task, such as
initialization, a more efficient scheme with complexity
\emph{O}($2^nn^2$) was proposed \cite{Long2}.

$C^n(U)$ gates are typical $n$-qubit fully controlled-$U$ gates that
apply a unitary $U$ to the target qubit if and only if all the first
$n-1$ control qubits are 1. Circuits for $C^2(U)$, $C^3(U)$ and
$C^4(U)$ gates have been constructed \cite{Sleator, Shende, Vatan,
Barenco}. But for the general case with $n\geq5$, the explicit
construction is absent. In this Letter, we present two different
construction schemes for an arbitrary $C^n(U)$ gate, one uses an
exponential and the other uses polynomial number of CNOT and
one-qubit gates. The polynomial complexity scheme is good for large
scale quantum computing. The exponential complexity scheme prevails
for a circuit with a small qubit number. In particular, they are
analytic. These results are very appealing in designing quantum
computer programming language, because it not only saves computing
time for its construction, but also avoids errors in numerical
construction because of error accumulation.

\emph{II. Exponential Construction Scheme}-- First we introduce some
notation. For a generic $n$-qubit circuit, its qubits are numbered
from the top  from 1 to $n$. $\wedge^{k}(V)$ stands for a
controlled-$V$ gate with $k$ control qubits and one target qubit, so
$C^n(U)$ gate is equally represented by $\wedge^{n-1}(U)$ whose
$n-1$ control qubits positioned at the top and the target qubit at
the bottom. Order of operations in an expression as well as in
circuits are performed from left to right.

\begin{figure}[here]
\includegraphics[width=6cm]{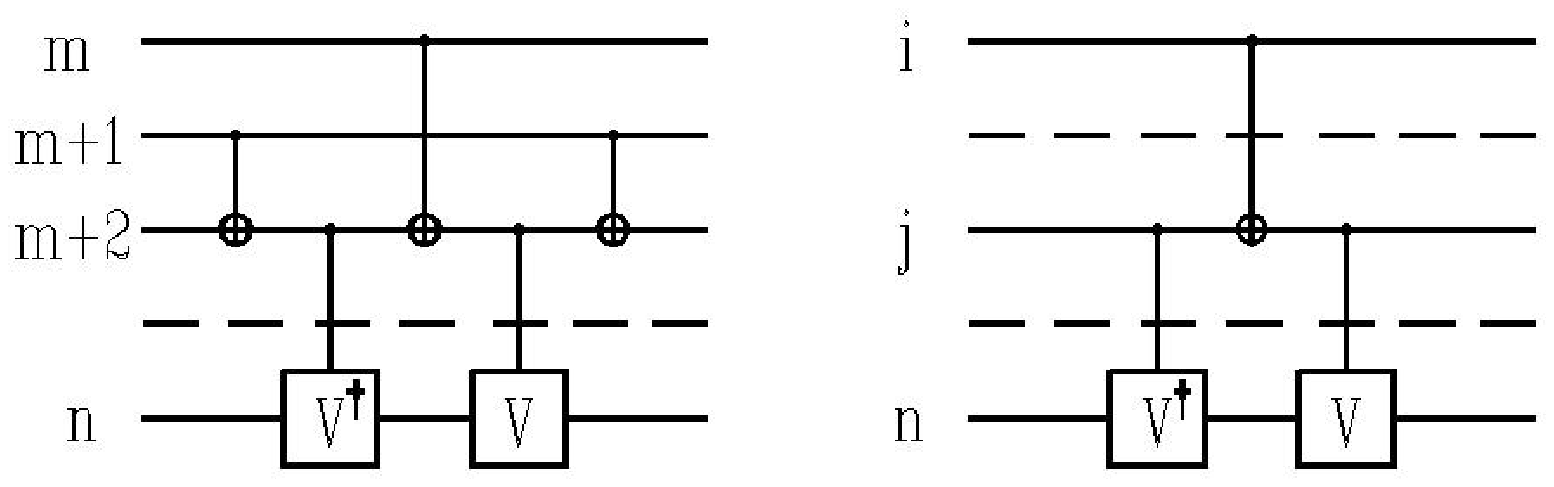}
\caption{ $A$ and $B$ blocks in repair section. The left part is a
$A$-block, the right one is a $B$-block.}\label{f1}
\end{figure}

Previous investigations gave explicit networks of $C^n(U)$ gates for
$n=2$, $3$, $4$. In this Letter, we present a general analytic
scheme implementing $C^n(U)$ gates for arbitrary values of $n$ and
any unitary operator $U$. Firstly, we define two kinds of quantum
gate-array blocks, the $A$-block and the $B$-block as shown in
Fig.\ref{f1}.

The $A$-block is indicated as $A^m$, where $m=1,\ldots,n-3$. Its
qubit nodes involve qubits $m$, $m+1$, $m+2$ and $n$. The $B$-block
is labeled as $B^i_j$, where $1\leq i< j< n$. Its qubit nodes
involve qubits $i$, $j$ and $n$. First we suppose the explicit
gate-array components of $C^{n-1}(U)$ network has been known, then
we give a general analytic expression. Our strategy for $C^n(U)$
network is a two-step procedure: basic section constructing in the
left part and repair section constructing in the right part of the
circuit. Basic section is obtained by combining $C^{n-1}(U)$ network
with a control input that is the $(n-1)$-th line without any
performance. The basic section of $C^n(U)$ network is indicated as
$\widetilde{C^{n-1}}$. $\widetilde{C^{n-1}}$ contains $2^{n-2}-2$
CNOT gates, $(2^{n-2}-1)$ number of $\wedge^1(V)$ and
$\wedge^1(V^\dag)$ gates, where $V^{2^{n-2}}=U$. Repair section is
yielded by placing $A^m$ and $B^i_j$ gate-array blocks in an
alternating sequence with respective number of $2^{n-4}$.

A $\beta$-bit Gray code \cite{Press} strings $\{g_\alpha\}$, where
$\alpha=1,\ldots,2^\beta$ is a palindromelike ordering with special
property that the adjacent bit strings differ only by a single bit.
We define a function $\gamma(\alpha,\beta)$ to represent the
numerical value of the position where $g_\alpha$ and $g_{\alpha+1}$
differ. In the repair section of $C^n(U)$, the index $m$ of $A^m$
block is definite as $n-3$, the index $j$ of $B^i_j$ blocks is
definite as $n-1$, whereas index $i$ varies complying with a
$(n-4)$-bit binary Gray code strings sequence. Denote
$\widetilde{C^{k}}$ as a network obtained from $C^{k}(U)$ gate
combined with $n-k$ extra qubits positioned between its last two
qubits. Carrying out the recursion, the following results are
obtained:
\begin{small}
\begin{eqnarray}
&&C^5(U)=\widetilde{C^4}A^2 B^1_4 A^2 B^1_4,\nonumber\\
&&C^6(U)=\widetilde{C^5}A^3 B^2_5 A^3 B^1_5 A^3 B^2_5 A^3 B^1_5,\nonumber\\
&&C^7(U)=\widetilde{C^6}A^4 B^3_6 A^4 B^2_6 A^4 B^3_6 A^4 B^1_6 A^4
B^3_6 A^4 B^2_6 A^4 B^3_6 A^4 B^1_6,\nonumber\\
&&\vdots\nonumber\\
&&C^n(U)=\widetilde{C^{n-1}}A^{n-3} B^{n-4}_{n-1} A^{n-3}
B^{n-3}_{n-1}\ldots A^{n-3} B^1_{n-1}.
\end{eqnarray}
\end{small}
So a generic $C^n(U)$ circuit where $n\geq 5$ can be expressed:
\begin{equation}
C^n(U)=\widetilde{C^4}\prod^{n-4}_{\beta=1}\prod^{2^\beta}_{\alpha=1}A^{\beta+1}B^{\gamma(\alpha,\beta)}_{\beta+3}.
\end{equation}
Given a unitary operator $V$, there must exist one-qubit unitary
operations $D$, $E$, $F$ and real number $a$ such that $DEF=I$ and
$e^{ia}D\sigma_x E\sigma_x F=V$. $\sigma_x$ and $G$ are unitary
one-qubit operations corresponding to matrices
$\sigma_x=\left[\begin{array}{cc}1&0\\0&1\end{array}\right]$ and
$G=\left[\begin{array}{cc}1&0\\0&e^{ia}\end{array}\right]$. Let
$C^k_{k^{\prime}}(U)$ denote a $\wedge^1(U)$ gate where qubit $k$
controls the qubit $k^{\prime}$. We rewrite Eq. (2) in terms of CNOT
and one-qubit gates after certain gate counting:
\begin{small}
\begin{eqnarray}
C^n(U)&=&\widetilde{C^4}\prod^{n-4}_{\beta=1}\prod^{2^\beta}_{\alpha=1}F^{\dag}_{n}C^{\beta+2}_{\beta+3}(\sigma_x)C^{\beta+3}_{n}(\sigma_x)G^{\dag}_{\beta+3}E^{\dag}_{n}C^{\beta+1}_{\beta+3}(\sigma_x)\nonumber\\
&&C^{\beta+1}_{n}(\sigma_x)G_{\beta+3}E_{n}C^{\beta+2}_{\beta+3}(\sigma_x)C^{\beta+2}_{n}(\sigma_x)G^{\dag}_{\beta+3}E^{\dag}_{n}\nonumber\\
&&C^{\gamma(\alpha,\beta)}_{\beta+3}(\sigma_x)C^{\gamma(\alpha,\beta)}_{n}(\sigma_x)G_{\beta+3}E_{n}C^{\beta+3}_{n}(\sigma_x)F_{n},\label{e3}
\end{eqnarray}
\end{small}
where
\begin{small}
\begin{eqnarray}
\widetilde{C^4}&=&D_{n}C^{1}_{n}(\sigma_x)E_{n}C^{2}_{n}(\sigma_x)C^{1}_{2}(\sigma_x)G^{\dag}_{2}E^{\dag}_{n}C^{2}_{n}(\sigma_x)C^{1}_{2}(\sigma_x)E_{n}\nonumber\\
&&C^{3}_{n}(\sigma_x)C^{2}_{3}(\sigma_x)E^{\dag}_{n}G^{\dag}_{2}C^{1}_{3}(\sigma_x)C^{1}_{4}(\sigma_x)G_{3}E_{n}C^{2}_{3}(\sigma_x)\nonumber\\
&&C^{2}_{n}(\sigma_x)G^{\dag}_{3}E^{\dag}_{n}C^{1}_{3}(\sigma_x)C^{1}_{n}(\sigma_x)E_{n}C^{3}_{n}(\sigma_x)G_{1}G_{2}G_{3}F_{n}.\nonumber\\
\label{e4}
\end{eqnarray}
\end{small}
In Eqs. (\ref{e3},\ref{e4}), $C^{k}_{k^{\prime}}(\sigma_x)$ are CNOT
gates and the $D$, $E$, $F$, $G$ and their hermitian conjugate are
the one-qubit gates and their subscripts represent the positions.
The $\beta$-bit binary Gray code strings reflected in a sequence of
$B^{\gamma(\alpha,\beta)}_{\beta+3}$ can be chosen freely in an
arbitrary cyclic $\beta$-qubit Gray code sequence and the $C^n(U)$
circuit for $(n\geq3)$ are self-inverse.

We can prove this exponential simulation fulfills the action of
$C^n(U)$ faithfully. After a carefully accounting of merges of CNOT
gates and single-qubit, we find this exponential simulation scheme
for a $C^{n}(U)$ gate finally utilizes $2^n-2$ CNOT gates and $2^n$
one-qubit gates.

\emph{III. Polynomial Construction Scheme} -- The above scheme is
advantageous for small values of $n$, but it becomes inefficient for
a large value of $n$, for instance $n>8$ because of its exponential
complexity. Here we propose a $C^{n}(U)$ circuit using
$\emph{O}(n^2)$ basic CNOT and one-qubit gates.

\begin{figure}[here]
\includegraphics[width=6cm]{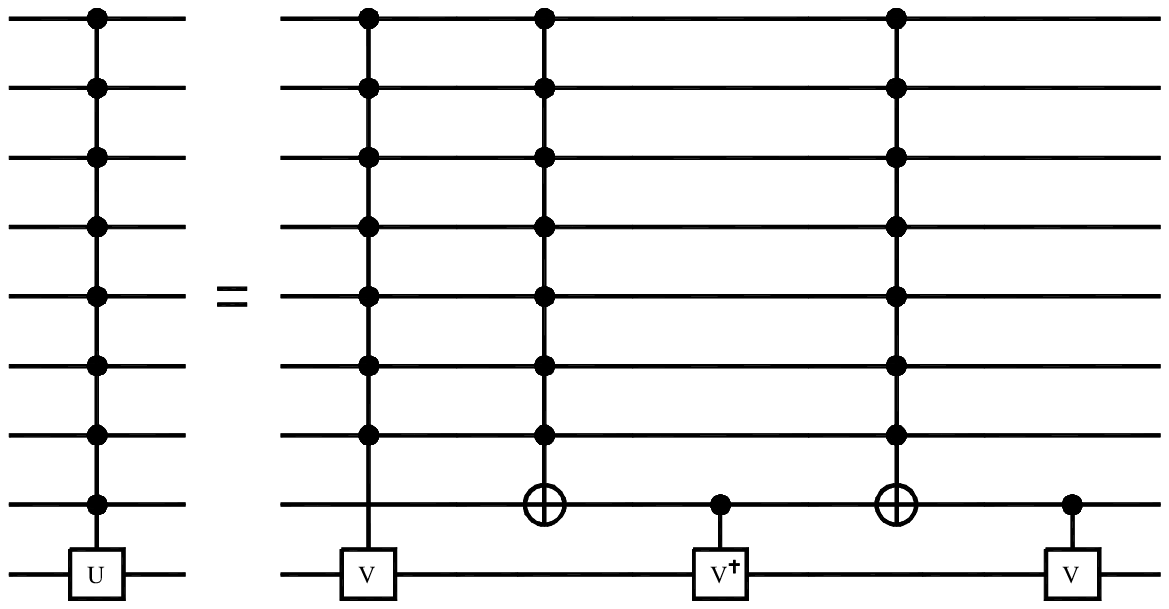}
\caption{A quantum circuit for implementing $C^n(U)$ gates with
unitary $V$ meeting $V^2=U$.}\label{f2}
\end{figure}

We know for $n\geq 3$, $C^{n}(U)$ can be simulated by a network with
its own inverse, where $V^2=U$ in Fig. \ref{f2}.

\begin{figure}[here]
\includegraphics[width=6cm]{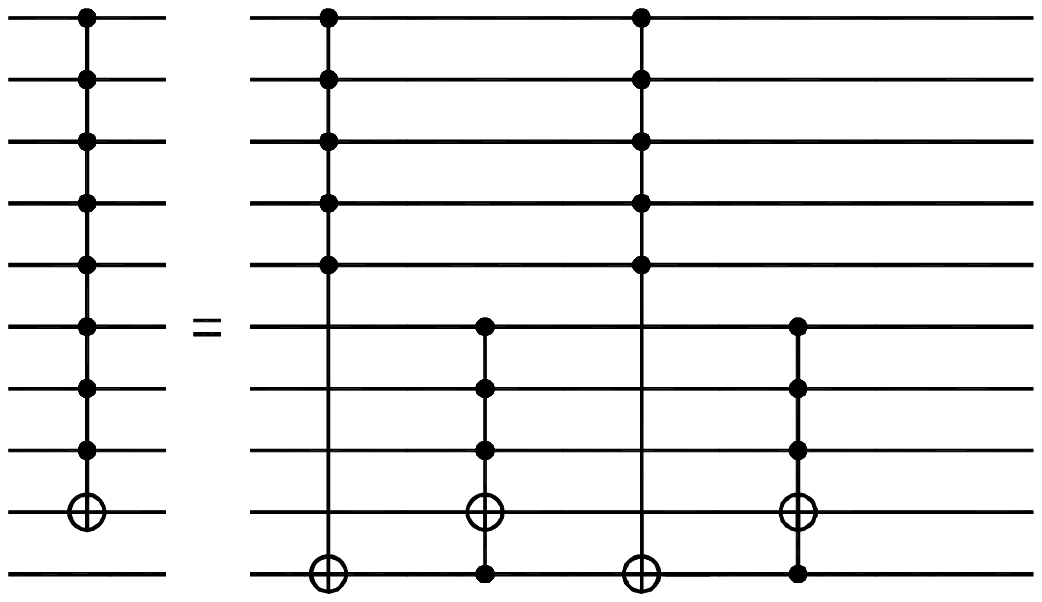}
\caption{A quantum circuit for implementing a
$\wedge^{n-2}(\sigma_x)$ gate with $\wedge^{m_1}(\sigma_x)$ and
$\wedge^{m_2}(\sigma_x)$ gates where $m_1+m_2=n-1$.}\label{f3}
\end{figure}

Given the explicit construction of arbitrary $C^{n-1}(U)$ gate is
known, the key procedure is to simulate two $\wedge^{n-2}(\sigma_x)$
gates. For $n\geq 4$ and $m_1\in {1,\ldots, n-2}$, a
$\wedge^{n-2}(\sigma_x)$ gate can be partitioned into two
$\wedge^{m_1}(\sigma_x)$ gates and two $\wedge^{m_2}(\sigma_x)$
gates, where $m_1+m_2=n-1$ as shown in Fig. \ref{f3}.

\begin{figure}[here]
\includegraphics[width=8cm]{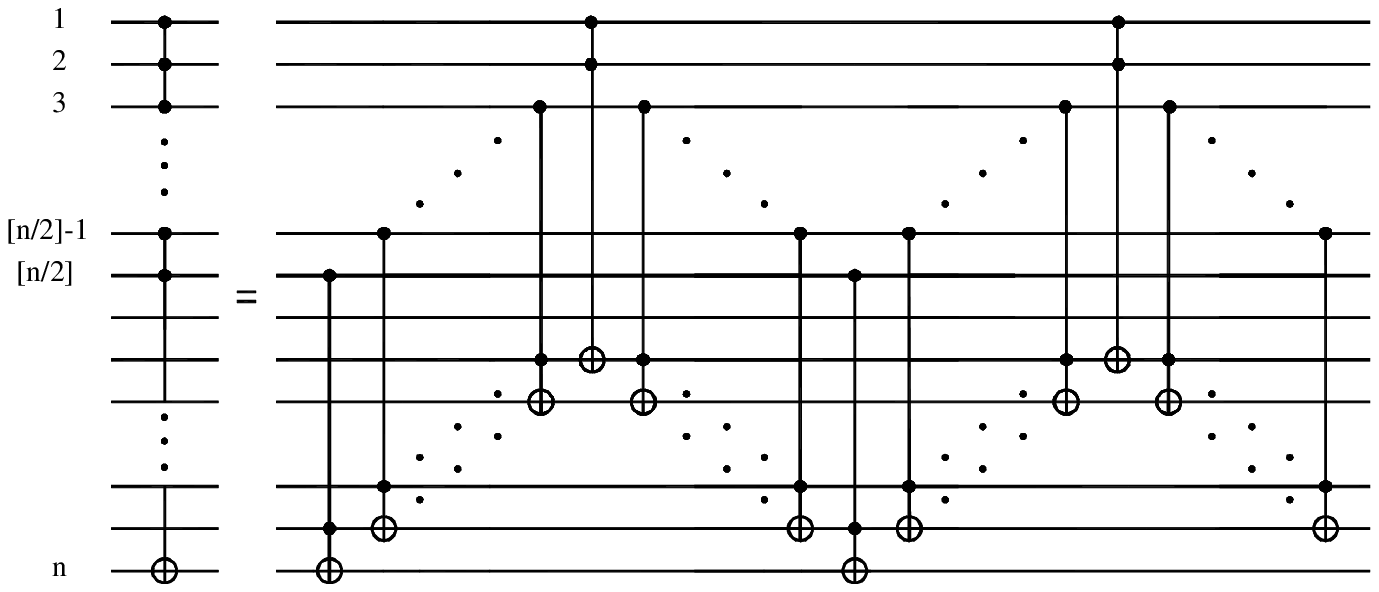}
\caption{A quantum circuit for a $\wedge^{m_1}(\sigma_x)$ gate
consisting of Toffoli gates arranged most regularly.}\label{f4}
\end{figure}

So the problem is reduced to how to construct
$\wedge^{m_1}(\sigma_x)$ and $\wedge^{m_2}(\sigma_x)$ gates. If we
assign $m_1=[n/2]$, $m_2=n-[n/2]-1$, $\wedge^{m_1}(\sigma_x)$ and
$\wedge^{m_2}(\sigma_x)$ can be decomposed into several Toffoli
gates. It is worthy noting that these decompositions are only
applicable to $\wedge^{m_1}(\sigma_x)$ for $n\geq 6$ and to
$\wedge^{m_2}(\sigma_x)$ for $n\geq 7$. So for $n\geq 6$, we
investigate the most regular arrangement of Toffoli gates
implementing $\wedge^{m_1}(\sigma_x)$ in Fig. \ref{f4}.
Let $T^{\substack{a\\b}}_c$ denote a Toffoli gate with control
qubits $a$ and $b$, the target qubit $c$. Consider the
characteristic features of such network: $\wedge^{m_1}(\sigma_x)$
network consists of $4[n/2]-8$ Toffoli gates and the indices of
Toffoli gates is symmetric  around $i_0=[n/2]-1$ and periodic with
period $d=2[n/2]-4$, then use a formula to describe the regulation
of Toffoli gates
\begin{equation}
\wedge^{m_1}(\sigma_x)=\prod^{4[n/2]-8}_{i=1}T^{\begin{subarray}{l}2+f(n)\\1+(1-\delta_{i,i_0})(1-\delta_{i,i_0+d})(n-[n/2]+f(n))\end{subarray}}_{n-[n/2]+2+f(n)},
\end{equation}
where
$f(n)=|\frac{d}{2}+\frac{d}{\pi}\arctan(\tan(\frac{\pi}{d}i-\frac{\pi}{2}))-[\frac{n}{2}]+1|$.
It denotes the deviation of $i$ to $\frac{n}{2}-1$ when $1\leq i\leq
d$ or to $\frac{n}{2}-1+d$ when ${d+1}\leq i\leq 2d$. The absolute
value function expresses symmetric property, the $\arctan$ function
fixes periodic regulations, and the $\delta$ functions correspond to
certain singular points at $i=[n/2]-1$ and $3[n/2]-5$ referred in
above formalism.

\begin{figure}[here]
\includegraphics[width=8cm]{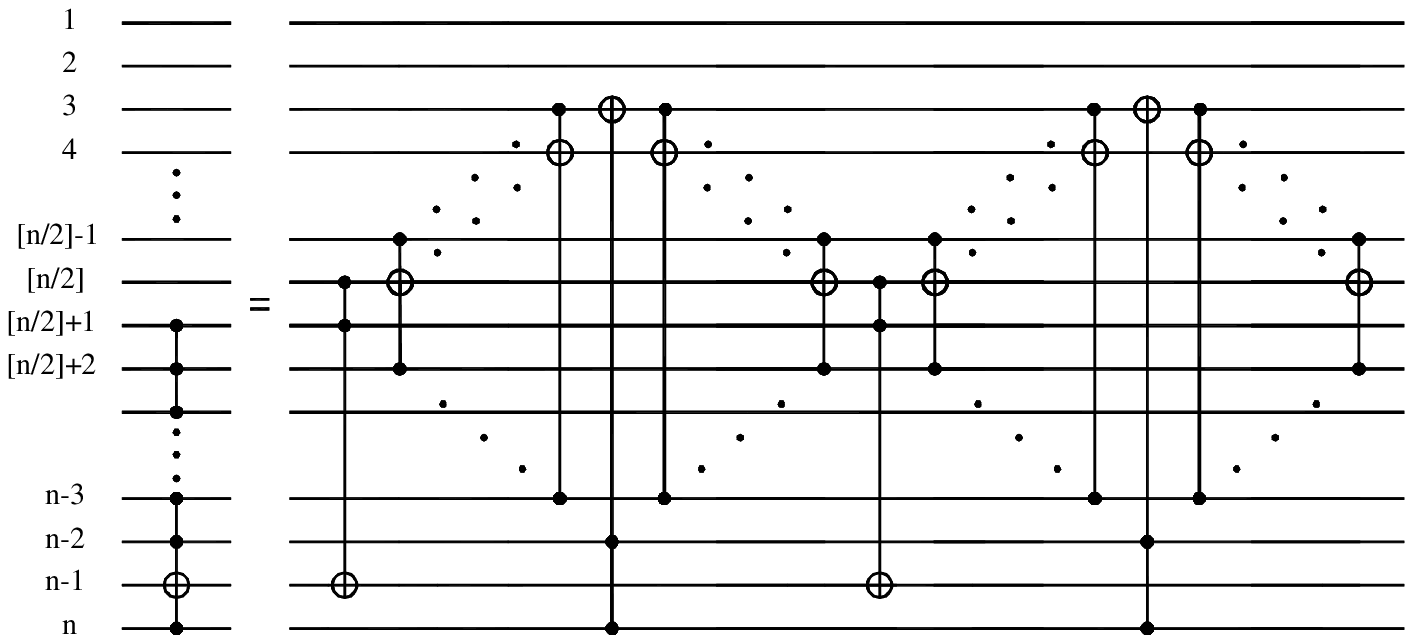}
\caption{A quantum circuit for a $\wedge^{m_2}(\sigma_x)$ gate
consisting of Toffoli gates arranged most regularly.}\label{f5}
\end{figure}

Similarly, for $n\geq 7$, $\wedge^{m_2}(\sigma_x)$ gates can be
simulated by the network in Fig. \ref{f5}.
We find that there are $4n-4[n/2]-12$ Toffoli gates in
$\wedge^{m_2}(\sigma_x)$ network. Inspecting the mathematic property
of indices, it is periodic with period $d'=2n-2[n/2]-6$, symmetric
around $j_0=n-[n/2]-2$ and $j_0+d'=3n-3[n/2]-8$, we obtain the
following formula
\begin{scriptsize}
\begin{eqnarray}
\wedge^{m_2}(\sigma_x)=\prod^{4n-4[n/2]-12}_{j=1}
T^{\begin{subarray}{l}n+(1-\delta_{j,j_0})(1-\delta_{j,j_0+d'})(2[n/2]-2n+3+g(n))\\n-2-g(n)\end{subarray}}_{n-1+(1-\delta_{j,1})(1-\delta_{j,2n-2[n/2]-5})(2[n/2]-2n+5+g(n))},
\end{eqnarray}
\end{scriptsize}
where
$g(n)=|\frac{d'}{2}+\frac{d'}{\pi}\arctan(\tan(\frac{\pi}{d'}j-\frac{\pi}{2}))-n+[\frac{n}{2}]+2|$.
Then we propose a cascade decomposition of $C^n(U)(n\geq 7)$ gate by
a recursive method shown in Fig. \ref{f6}, where unitary $V_i$ is
defined by $V^{2^i}_i=U$.
\begin{widetext}
\begin{center}
\begin{figure}[here]
\includegraphics[width=14cm]{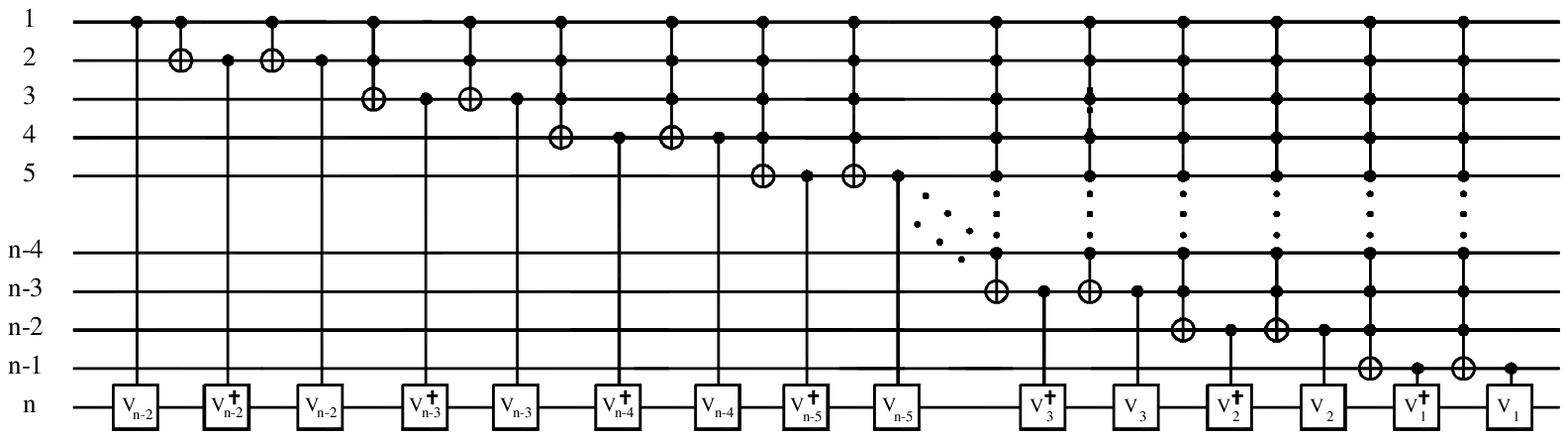}
\caption{Cascade structure of $C^n(U)$ gates where unitary $V_i$
satisfies $V^{2^i}_i=U$.}\label{f6}
\end{figure}
\end{center}
\end{widetext}

\begin{figure}[here]
\includegraphics[width=8cm]{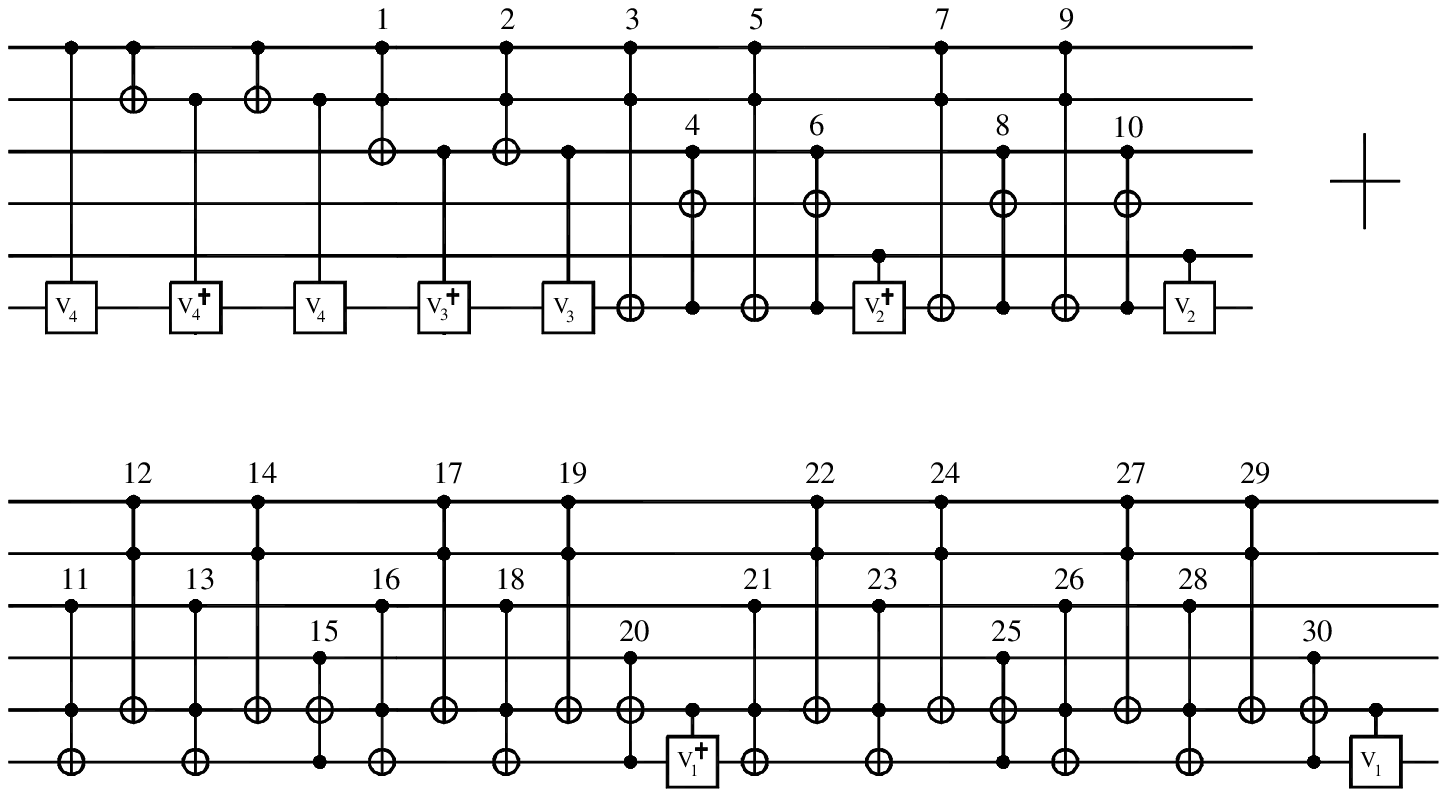}
\caption{The explicit structure of $C^6(U)$ (congruent to
$\widetilde{C^6(V_{n-6})}$)  in terms of Toffoli  and two-qubit
controlled gates.}\label{f7}
\end{figure}

Suppose $T^{\{1,2,\ldots k-1\}}_k$ is a $\wedge^{k-1}(\sigma_x)$
whose first $k-1$ qubits control the last one qubit. Then above
decomposition can be described in formula
\begin{tiny}
\begin{eqnarray}
C^n(U)=\widetilde{C^{n-1}(V_1)}T^{\{1,2,\ldots
n-2\}}_{n-1}C^{n-1}_n(V^\dag_1)T^{\{1,2,\ldots
n-2\}}_{n-1}C^{n-1}_n(V_1)&&\nonumber\\
=\widetilde{C^6(V_{n-6})}\prod^n_{k=7}T^{\{1,2,\ldots
k-2\}}_{k-1}C^{k-1}_n(V^\dag_{n-k+1})T^{\{1,2,\ldots
k-2\}}_{k-1}C^{k-1}_n(V_{n-k+1}),&&\nonumber\\
&&
\end{eqnarray}
\end{tiny}
where $T^{\{1,2,\ldots
k-2\}}_{k-1}=\wedge^{m_1}(\sigma_x)\wedge^{m_2}(\sigma_x)\wedge^{m_1}(\sigma_x)\wedge^{m_2}(\sigma_x)$.
After tedious calculation, we find out a $C^n(U)$ gate totally
requires 2 CNOT gates, $8n^2-72n+174$ Toffoli gates and $2n-3$
two-qubit controlled gates. Then the problem is reduced to
simulating Toffoli gate with basic CNOT and one-qubit gates. Using
well-known congruent modulo phase shift (CMPS) methods
\cite{DiVincenzo} for Toffoli gates, it can be expressed as
$T^{\begin{subarray}{l}a\\{b}\end{subarray}}_{c}=R_cC^b_c
(\sigma_x)R_cC^a_c(\sigma_x)R^\dag_cC^b_c(\sigma_x)R^\dag_c$, where
$R=R_y(\pi/4)$. The CMPS scheme only requires 7 basic operations
which is much less than 14 basic operations in the usual simulation
scheme. $\widetilde{C^6(V_{n-6})}$ part is congruent to the circuit
for $C^6(U)$ and we have proven that the Toffoli gates labeled as 4,
6, 8, 10, 15, 20, 25, 30 as shown in Fig. \ref{f7} for
$\widetilde{C^6(V_{n-6})}$ part, and all the Toffoli gates other
than $\widetilde{C^6(V_{n-6})}$ in Eq. (7) can be replaced by the
modulo phase shift of Tollofi gates.

Given a unitary operator  $V_{n-k+1}$, $L$, $P$, $R$ and $S$ are
one-qubit unitary gates such that
$e^{ib}L\sigma_xP\sigma_xQ=V_{n-k+1}$, $LPQ=I$ and
$S=\left[\begin{array}{cc}1&0\\0&e^{ib}\end{array}\right]$, their
subscripts represent which qubit they are performed on. Now we
obtain the $C^n(U)$ in terms of CNOT and one-qubit gates:
\begin{footnotesize}
\begin{eqnarray}
C^n(U)&=&\widetilde{C^6(V_{n-6})}\prod^n_{k=7}W_{k-1}W_{k-1}S^{\dag}_{k-1}Q^{\dag}_{n}C^{k-1}_{n}(\sigma)P^{\dag}_{n}C^{k-1}_{n}(\sigma)L^{\dag}_{n}\nonumber\\
&&W_{k-1}W_{k-1}L_{n}C^{k-1}_{n}(\sigma)P_{n}C^{k-1}_{n}(\sigma)S_{k-1}Q_{n},\nonumber\\
\end{eqnarray}
\end{footnotesize}
where
\begin{small}
\begin{eqnarray}
W_{k-1}=&&\nonumber\\
\{\prod^{4[k/2]-8}_{i=1}&&R_{k-[k/2]+2+f(k)}\nonumber\\
&&C^{1+(1-\delta_{i,i_0})(1-\delta_{i,i_0+d})(k-[k/2]+f(k))}_{k-[k/2]+2+f(k)}(\sigma_x)\nonumber\\
&&R_{k-[k/2]+2+f(k)}C^{2+f(k)}_{k-[k/2]+2+f(k)}(\sigma_x)R^{\dag}_{k-[k/2]+2+f(k)}\nonumber\\
&&C^{1+(1-\delta_{i,i_0})(1-\delta_{i,i_0+d})(k-[k/2]+f(k))}_{k-[k/2]+2+f(k)}(\sigma_x)\nonumber\\
&&R^{\dag}_{k-[k/2]+2+f(k)}\}\nonumber\\
\{\prod^{4k-4[k/2]-12}_{j=1}&&R_{k-1+(1-\delta_{j,1})(1-\delta_{j,1+d'})(2[k/2]-2k+5+g(k))}\nonumber\\
&&C^{k-2-g(k)}_{k-1+(1-\delta_{j,1})(1-\delta_{j,1+d'})(2[k/2]-2k+5+g(k))}(\sigma_x)\nonumber\\
&&R_{k-1+(1-\delta_{j,1})(1-\delta_{j,1+d'})(2[k/2]-2k+5+g(k))}\nonumber\\
&&C^{k+(1-\delta_{j,j_0})(1-\delta_{j,j_0+d'})(2[k/2]-2k+3+g(k))}_{k-1+(1-\delta_{j,1})(1-\delta_{j,1+d'})(2[k/2]-2k+5+g(k))}(\sigma_x)\nonumber\\
&&R^{\dag}_{k-1+(1-\delta_{j,1})(1-\delta_{j,1+d'})(2[k/2]-2k+5+g(k))}\nonumber\\
&&C^{k-2-g(k)}_{k-1+(1-\delta_{j,1})(1-\delta_{j,1+d'})(2[k/2]-2k+5+g(k))}(\sigma_x)\nonumber\\
&&R^{\dag}_{k-1+(1-\delta_{j,1})(1-\delta_{j,1+d'})(2[k/2]-2k+5+g(k))}\}.
\end{eqnarray}
\end{small}

Taking account of the merges of CNOT gates and one-qubit gates, we
obtain the total number of basic operations in $C^n(U)$ construction
are $24n^2-212n+540$ CNOT gates and $32n^2-288n+739$ one bit gates
ultimately.

\emph{IV. Conclusions}-- In conclusion, we have given two analytic
schemes for constructing a $C^n(U)$ gate for arbitrary value of $n$
and any unitary $U$ operator, one with exponential complexity and
the other with polynomial complexity. General expression for
decomposition of $C^n(U)$ gats with basic one-qubit gates and CNOT
gates has been derived explicitly. We have compared the exact
numbers of basic operations required in these two methods for
$n=1-20$. It shows that the exponential construction is advantageous
for the value of $n=1-8$, whereas the polynomial simulation is
efficient for larger values of $n>8$.

This work is supported by the National Fundamental Research Program
Grant No. 2006CB921106, China National Natural Science Foundation
Grant Nos. 10325521, 60433050.

\end{document}